\documentstyle[11pt,AATS,epsf]{article}	

\newcommand{\hbeta}{H$\beta$}
\newcommand{\hga}{H$\gamma_A$}

\begin{document}	

\title{The Formation Timescales of Giant Spheroids} 

\author{S. C. Trager}
\affil{The Observatories of the Carnegie Institution of Washington,
813 Santa Barbara Street, Pasadena, CA 91101}
\author{S. M. Faber}
\affil{UCO/Lick Observatory, Department of Astronomy \& Astrophysics,
University of California, Santa Cruz, CA 95064}
\author{A. Dressler}
\affil{The Observatories of the Carnegie Institution of Washington,
813 Santa Barbara Street, Pasadena, CA 91101}


\begin{abstract}
We review current progress in the study of the stellar populations of
early-type galaxies, both locally and at intermediate redshifts.  In
particular, we focus on the ages of these galaxies and their evolution
in hopes of determining the star formation epochs of their stars.  Due
to serious remaining systematic uncertainties, we are unable to
constrain these epochs precisely.  We discuss our results on the
evolution of stellar populations in the context of other observables,
in particular the evolution of the Fundamental Plane of early-type
galaxies.
\end{abstract}

\section{Introduction}

The sequence and timescales of galaxy formation are a major issue in
astrophysics today.  The formation histories of early-type galaxies,
ellipticals and S0's, though seemingly simple systems, remain a
puzzle.  Long thought to be very old, homogeneous, coeval systems
varying only in their metallicities (Baade 1962; Faber 1977; Burstein
1977), accumulating observational evidence on stellar populations of
ellipticals and S0's have challenged this view.  A number of field
elliptical galaxies clearly have had significant recent star formation
(see below), and many early-type galaxies have evidence for recent
dynamical disturbances (Schweizer \& Seitzer 1992).  Theoretically,
models of the commonly accepted mode of galaxy
formation---hierarchical clustering (e.g.\ Blumenthal et al.\ 1984;
Kauffmann, White, \& Gui\-der\-do\-ni 1993)---suggest that merging,
accretion, and star formation have continued in the giant early-type
galaxy population up until the current epoch, at least in some
galaxies in low-density environments.  However, these models are
unconstrained by detailed observations of the formation epoch
(parameterized by the formation redshift $z_f$ of the dominant stellar
population) and the timescale of star formation in early-type
galaxies.

Detailed studies of the stellar populations of early-type galaxies
allow us to determine or at least infer these timescales.  In
particular, ages derived from stellar populations and their evolution
with redshift allow us to determine \emph{when} the stars in giant
spheroids form, and their nucleosynthetic properties (their
metallicities and relative abundances such as [$\alpha$/Fe]) allow us
to determine \emph{how long} the star formation event(s) lasted.

Unfortunately, such detailed data are difficult to gather due to the
\emph{age-metallicity degeneracy} (e.g., O'Connell 1986; Worthey 1994;
Trager 1999), which plagues all stellar population studies at some
level.  As stellar populations age, the populations get cooler and
thus redder.  Unfortunately, stellar populations also get cooler and
redder with increasing metallicity.  Colors and metal absorption lines
like Mg$_2$ are therefore degenerate to compensating changes in age
and metallicity.  Worthey (1994, following earlier work by O'Connell
1980, Rabin 1982, Burstein et al.\ 1984, and Rose 1985) showed that
the Balmer lines of hydrogen can break this age-metallicity
degeneracy.  These absorption lines originate in the hot main-sequence
turnoff (MSTO) stars, and the MSTO temperature is much more sensitive
to age than to metallicity in stellar populations (cf.\ Fig.\ 1 of
Trager 1999).  However, because even the Balmer lines are still
somewhat sensitive to metallicity, a combination of Balmer and metal
lines are used to determine ages and abundances of early-type galaxies
(Figs.~\ref{fig:local} and \ref{fig:distant}; e.g., Gonz\'alez 1993;
Worthey 1994; J{\o}rgensen 1997, 1999, this meeting; Kuntschner 2000;
Trager et al.\ 2000a,b).

In this contribution, we summarize the current state of our own
studies of the stellar populations of both local and distant
early-type galaxies.  We focus on our attempts to determine the
typical formation redshifts of these galaxies in various environments
and on the difficulties in measuring---and comparing---the stellar
populations of early-type galaxies.  

Throughout this contribution, we use the stellar population models of
Wor\-they (1994) for consistency with earlier work, but note that the
absolute time\-scale of these models (that is, MSTO temperature at
fixed age) is suspect, with the oldest models being too old by about
25--35\% when compared with models based on recent isochrones by the
Padova group (e.g., Girardi et al.\ 2000; cf.\ Charlot, Worthey \&
Bressan 1996).  However, the \emph{relative} ages of galaxies are
nearly unaffected by the choice of stellar population model (Trager et
al.\ 2000a), which are the data of interest here.  Also, it is helpful
to keep in mind that the ages (and abundances) presented here are
those of single-burst, single-metallicity stellar populations (SSPs);
see Trager et al.\ (2000b) for details of the effects of composite
populations on SSP parameters.

\section{The Stellar Populations of Nearby Early-Type Galaxies}

\begin{figure}
\plotsetheight{6cm}{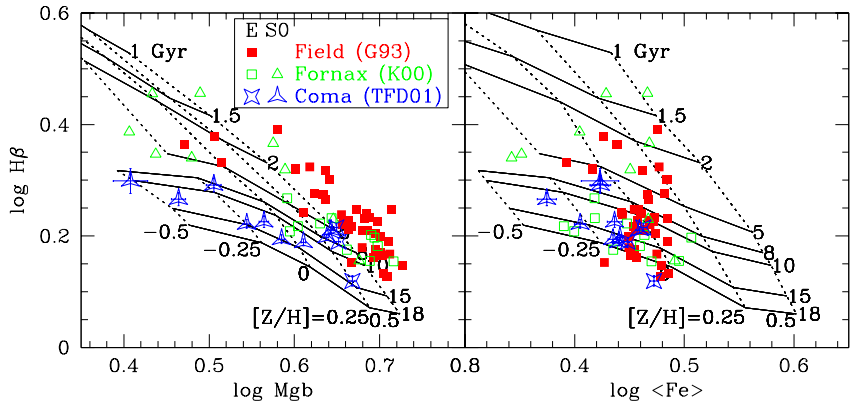}
\caption{The line strengths of local early-type galaxies, by
morphology and environment.  Squares are ellipticals, triangles are
S0's, solid points are ``field'' (isolated, group, and Virgo cluster)
galaxies from Gonz\'alez (1993), open points are Fornax Cluster
galaxies from Kuntschner (2000), and stellated points are Coma Cluster
galaxies (Trager et al., in prep.).  Note the large dispersion in the
\hbeta\ strengths---and thus ages---of field ellipticals and Fornax
S0's but the rather tight distribution in the \hbeta\ strengths of
Fornax ellipticals and Coma galaxies, both ellipticals and
S0's.\label{fig:local}}
\end{figure}

At the present, only a limited number of early-type galaxies (less
than 100 total ellitpicals and S0's) have had absorption line
strengths measured accurately enough for detailed stellar population
work (that is, $\sigma_{H\beta}<0.05$ \AA, or S/N $>75$\AA).  Many of
these galaxies are shown in Figure~\ref{fig:local}, in which the
galaxies are separated by morphology (elliptical vs.\ S0) and by
environment (cluster vs.\ field).  A full analysis of the stellar
populations of the field and Fornax Cluster ellipticals is given by
Trager et al.\ (2000b), but the salient points for a discussion of the
formation epochs of early-type galaxies can be gleaned directly from
this figure.

\begin{description}
\item[Field ellipticals.] Field ellipticals from Gonz\'alez (1993),
which include ellipticals in environments ranging from isolated
galaxies to Virgo Cluster galaxies, show a wide spread in age.  This
spread in age should not be interpreted necessarily as indicative of a
wide range of formation redshifts for the entire stellar populations
of these object.  Indeed, Trager et al.\ (2000b) have interpreted this
spread in the context of two-burst models of star formation: Small
``frostings'' of recent star formation occur on top of massive old
stellar populations, with the range in inferred ages corresponding to
some combination of the time at which these frostings were formed and
the strength of the burst.
\item[Cluster ellipticals.] Cluster ellipticals in Virgo (Gonz\'alez
1993; Trager et al.\ 2000b), Fornax (Kuntschner 2000), and in the core
of Coma (Trager, Faber, \& Dressler, in prep.)\ appear to be coeval to
within a few Gyr, with virtually no recent star formation, except in a
few Virgo galaxies.
This lack of detectable recent star formation is in agreement with
color-magnitude studies of early-type cluster galaxies (e.g., Bower,
Lucey, \& Ellis 1992).
\item[Cluster S0's.] Any distinction between field and cluster S0's is
less clearly defined than for ellipticals.  The stellar populations of
Fornax S0's (Kunt\-schner 2000) certainly span a large range in age
(similar to the stellar populations of field S0's; Fisher, Franx, \&
Illingworth 1996), but those in the core of Coma (Trager et al., in
prep.)\ appear to have a rather tight age distribution, similar to the
cluster ellipticals.  A detection of any possible environmental effect
in the stellar populations of local S0's will require a much larger
database of high-quality spectra than is currently available.
\end{description}

\section{The Evolution of the Stellar Populations of Cluster Galaxies}

The lack of detectable recent star formation in the majority of
cluster ellipticals suggests that such galaxies can be used as direct
tracers of the evolution of the oldest stellar populations.  In this
section, we present the first results of our on-going study of the
evolution of the stellar populations of cluster galaxies (Trager 1997;
Trager, Faber, \& Dressler, in prep.).

Many techniques are available to detect and characterize the evolution
of early-type galaxies.  Here we concentrate on two methods: direct
detection of the evolution of stellar population ages and the
evolution of line strength--velocity dispersion relations.  (Other
methods, such as the evolution of cluster color-magnitude diagrams and
the evolution of the morphology-density relation, are described
elsewhere; see, e.g., Lubin, this volume).  We compare these two
methods with the evolution of the Fundamental Plane of early-type
galaxies in the next section.

\begin{figure}
\plotsetheight{6cm}{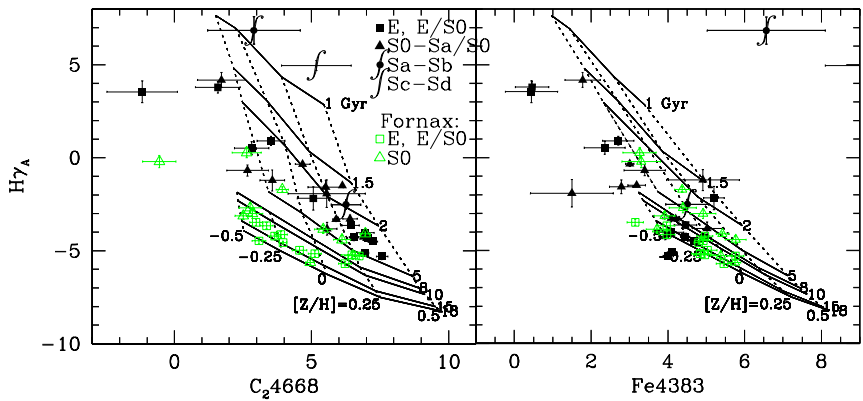}
\caption{The evolution of the absorption line strengths of cluster
galaxies.  Galaxies are coded by redshift (open points are Fornax
Cluster galaxies at $z\approx0$; solid points are galaxies in Abell
851 at $z=0.41$) and morphology (squares are ellipticals and E/S0
transition cases; triangles are S0's, S0/Sa, and Sa/S0 transition
cases; later-type galaxies are integral signs).  Evolution is detected
at the $5\sigma$ level in the \hbeta\ strengths of early-type cluster
galaxies, modulo uncertainties in the corrections needed to bring the
Fornax galaxies to the same very large physical apertures as the
distant galaxies and line strength system calibrations in the distant
galaxies (see text).\label{fig:distant}}
\end{figure}

We begin with the most ambitious method, direct detection of the
evolution of stellar populations ages of early-type galaxies.
Figure~\ref{fig:distant} shows our first attempt: a comparison of line
strengths of early-type galaxies in Abell 851 (=CL0939+4713), a rich
cluster at $z=0.406$, to those in Fornax.  Focussing first on the
elliptical galaxies (squares), the ellipticals in Abell 851 are about
40\% the age of the Fornax ellipticals ($>5\sigma$ significance).
However, this result is subject to three major uncertainties. (1) The
distant galaxies are not necessarily on the same line strength system
as the local galaxies; in particular, the C$_2$4668 line strengths may
be systematically uncertain by up to 0.25 \AA, and there may be
overall calibration issues with the poorly understood \hga\ index.
(2) The model line strengths of \hga\ appear to be too weak by about 1
\AA\ on comparison with stellar population parameters derived from
\hbeta\ for Fornax galaxies.  This offset appears to be due to
uncertainties in the emipirically-determined fitting functions
(Worthey \& Ottaviani 1997).  While this offset has been corrected in
Fig.~\ref{fig:distant}, its exact magnitude is still uncertain.  (3)
Because all early-type galaxies have gradients in their line strengths
(e.g., Davies, Sadler, \& Peletier 1993), galaxies must be measured
through the same physical aperture in order to directly compare their
line strengths and thus their stellar populations.  Unfortunately, the
extraction aperture used for the distant galaxies
($1\arcsec\times2\farcs4$) projects to a very large aperture on the
Fornax galaxies ($\sim1\arcmin\times2\arcmin$), much larger than the
apertures through which accurate line strengths for local galaxies
have been measured (a result anticipated by Kennicutt 1992).  We are
therefore forced to apply to the rather uncertain gradients of
early-type galaxies to generate aperture corrections for the
\emph{local} galaxies.  This is currently our largest uncertainty;
high S/N, raster-scanned spectroscopy of local early-type galaxies
(cf.\ Kennicutt 1992) will be required to resolve this issue.

\begin{figure}
\plotone{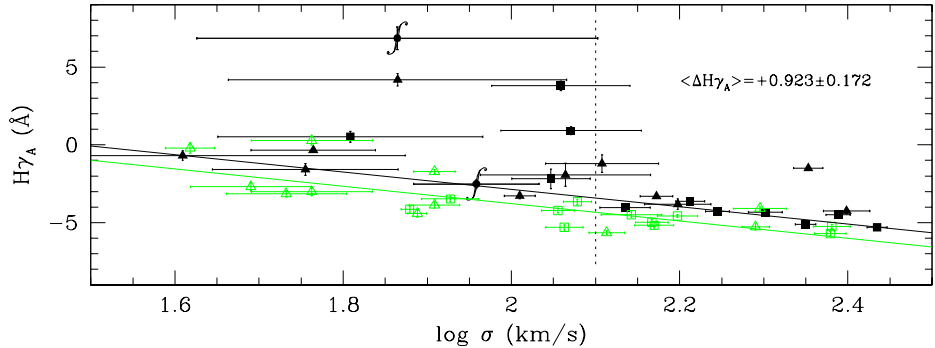}
\caption{The evolution of the Balmer line--velocity dispersion
relation (cf.\ Kelson et al.\ 2001).  Assuming all the evolution in
\hga\ is due to age, galaxies at $z\approx0.4$ are 70\% the age of
Fornax galaxies of the same velocity dispersion (for typical stellar
population models).
\label{fig:hgasigma}}
\end{figure}

Another method is to use the evolution of line strength--velocity
dispersion relations.  Although the pioneering work of Bender,
Ziegler, \& Bruzual (1996) used the Mg b--$\sigma$ relation, using
more age-sensitive indices like the high-order Balmer lines (Kelson et
al.\ 2001) provides much better leverage on the evolution.
Figure~\ref{fig:hgasigma} shows the evolution of the \hga--$\sigma$
relation from $z=0.41$ (Abell 851) to $z\approx0$ (Fornax).  Assuming
that the evolution is \emph{entirely due to age evolution at fixed
velocity dispersion}---and not due to metallicity differences between
cluster galaxies of the same velocity dispersion in the two
clusters, to changes in the \emph{slope} of the relation (if, say, the
lower-mass galaxies are younger than the older galaxies; cf.\ Trager
et al.\ 2000b), or some other manifestation of the age-metallicity
degerneracy---then the galaxies in Abell 851 are 70\% the age of the
Fornax galaxies at fixed velocity dispersion using common stellar
population models.  This result has the same systematic uncertainties
as the previous method---aperture corrections, line strength
calibrations, and model uncertainties---and moreover, since only a
single absorption line is used, the age-metallicity degeneracy can
play a significant role.

\section{Discussion and Conclusions}

\begin{figure}
\plotsetheight{11cm}{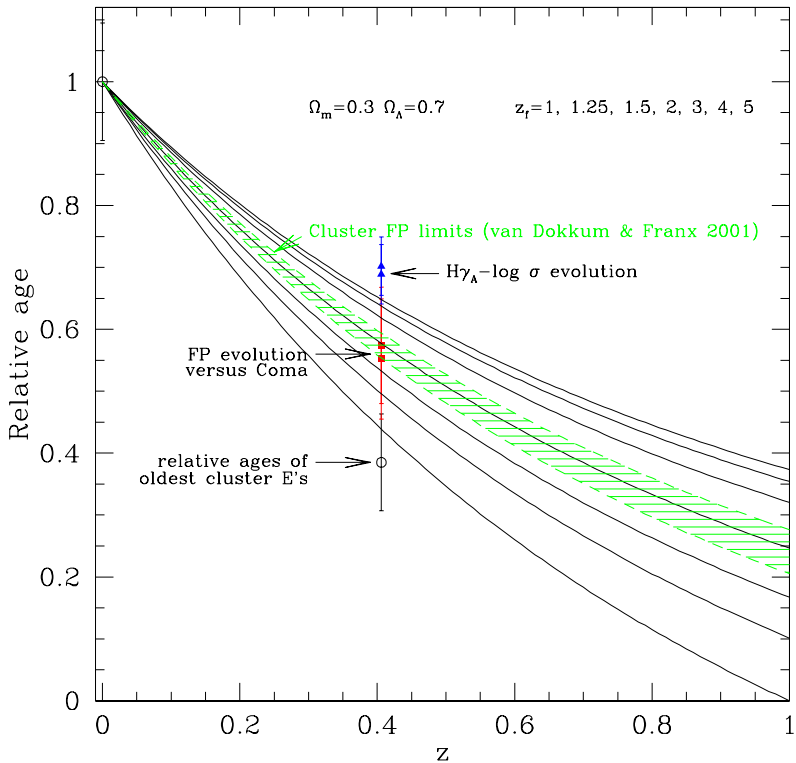}
\caption{Constraints on the formation redshift of early-type cluster
galaxies from three stellar population evolutionary indicators.  The
open circle represents the limits from the relative ages of the oldest
cluster ellipticals in Fornax ($z=0$) and Abell 851 ($z=0.41$) from
Fig.\ 2; the solid triangles represent the limits from the evolution
of the \hga--$\sigma$ relation from Fornax to Abell 851 from Fig.\ 3
(two different stellar population models); and the solid squares
represent the evolution of the Fundamental Plane from Coma to Abell
851 (photometry from Ziegler et al.\ 1999; authors' own velocity
dispersions).  The solid lines represent passively evolving stellar
populations with formation redshifts $z_f=1$, 1.25, 1.5, 2, 3, 4, 5,
from bottom to top.  The hatched region is the best fit from the
evolution of the FP from $z\approx0$ to $z=0.83$, including a
correction based model of morphological transformations onto the FP
(van Dokkum \& Franx 2001).  Due to uncertainties in aperture
corrections and line strength system calibrations in Figs.\ 2 and 3
and the age-metallicitiy degeneracy in Figs.\ 3 and 4, the formation
redshift of early-type cluster galaxies is still ill-constrained,
although $1\la z_f\la5$ is certainly reasonable, with $z_f\approx2$
preferred by the FP studies.
\label{fig:summary}}
\end{figure}

We summarize the results from these methods and the evolution of the
Fundamental Plane (van Dokkum \& Franx 2001; Trager et al.\ in prep.)\
in Figure~\ref{fig:summary}; at the moment, these three methods are
marginally inconsistent.  We have tried to point out the various
difficulties involved in measuring these evolutionary indicators: the
line strengths required extremely high signal-to-noise and therefore
efficient spectrographs on large telescopes, and systematic
uncertainties may be large.  Systematic uncertainties aside, another
effect may be present.  Small amounts of recent star formation can
strengthen the Balmer lines far out of proportion to the actual mass
involved in the burst (Trager et al.\ 2000b).  A recent study of
NUV-optical colors of galaxies in Abell 851 (Ferreras \& Silk 2000)
suggests that such recent star formation may have occured in many of
these objects.  This might explain the enhanced \hga\ strengths of the
distant galaxies without constraints from the evolution of the
Fundamental Plane of early-type galaxies (e.g., van Dokkum \& Franx
2001), but this explanation appears to be inconsistent with the
evolution of the \hga--$\sigma$ relation.

In conclusion, we have \emph{directly detected evolution} in the
stellar populations of early-type cluster galaxies out to a redshift
of $z=0.41$.  However, due to systematic uncertainties in measuring
and comparing absorption line strengths, the exact amount of the
evolution and the significance of this result is still unknown.
Formation redshifts of $1\la z_f\la5$ are consistent with the current
data (Fig.~\ref{fig:summary}), but a more precise answer awaits better
calibration of both the distant \emph{and} local data.

\acknowledgments

SCT is grateful to D. Kelson for code and welcome criticism, to
F. Schweizer for helpful conversations, to the organizers for a truly
enjoyable meeting, and to the editors for their patience and
understanding in allowing us to present a slightly more pessimistic
review than the one delivered at the meeting.  Support for this work
was provided by NASA through Hubble Fellowship grant HF-01125.01-99A
to SCT awarded by the Space Telescope Science Institute, which is
operated by the Association of Universities for Research in Astronomy,
Inc., for NASA under contract NAS 5-26555; by a Starr Fellowship to
SCT; by a Carnegie Fellowship to SCT; by a Flintridge Foundation
Fellowship to SCT; by NSF grant AST-9529098 to SMF; and by NASA
contract NAS5-1661 to the WF/PC-I IDT.

\end{document}